\documentclass[prl,aps,epsfig,superscriptaddress,showpacs]{revtex4}
\usepackage{epsfig,amsmath}
\begin{document}
\title{Gisin's Theorem for Three Qubits}
\author{Jing-Ling Chen}
\email{phycj@nus.edu.sg}
\affiliation{Department of Physics, National University of
Singapore, 2 Science Drive 3, Singapore 117542}
\affiliation{Institute of Applied Physics and Computational Mathematics,
P.O. Box 8009 (26), 100088 Beijing, China}
\author{Chunfeng Wu}
\affiliation{Department of Physics, National University of
Singapore, 2 Science Drive 3, Singapore 117542}
\author{L. C. Kwek}
\affiliation{Department of Physics, National University of
Singapore, 2 Science Drive 3, Singapore 117542}
\affiliation{Nanyang Technological University, National Institute
of Education, 1, Nanyang Walk, Singapore 637616}
\author{C. H. Oh}
\email{phyohch@nus.edu.sg}
\affiliation{Department of Physics, National University of
Singapore, 2 Science Drive 3, Singapore 117542}

\begin{abstract}
We present a Theorem that all generalized Greenberger-Horne-Zeilinger states of a three-qubit system violate a Bell inequality in terms of
probabilities. All pure entangled states of a three-qubit system are shown to violate a Bell inequality for
probabilities; thus, one has Gisin's theorem for three qubits.
\end{abstract}
\maketitle

Quantum mechanics violates
Bell type inequalities that hold for any local-realistic theory
\cite{bell,chsh,ghz,mermin,jlch}.
In 1991, Gisin presented a theorem, which states that {\it any}
pure entangled state of two particles violates a Bell inequality
for two-particle correlation functions \cite{gisin,popescu}.
Bell's inequalities for
systems of more than two qubits are the object of renewed
interest, motivated by the fact that entanglement between more
than two quantum systems is becoming experimentally feasible.
Recent investigations show the surprising result that there exists a
family of pure entangled $N>2$ qubit states that do not violate
any Bell inequality for $N$-particle correlations for the case of
a standard Bell experiment on $N$ qubits \cite{scarani}. By a standard Bell
experiment we mean the one in which each local observer is given a choice
between two dichotomic observables
\cite{zukowski2,zukowski1,weinfurter,werner}. This family is the generalized
Greenberger-Horne-Zeilinger (GHZ) states given by
\begin{eqnarray}
|\psi\rangle_{\rm GHZ}=\cos\xi|0\cdots 0\rangle+\sin\xi|1\cdots 1\rangle
 \label{eq1}
\end{eqnarray}
with $0 \le \xi \le \pi/4$. The GHZ states \cite{ghz} are for $\xi=\pi/4$. In 2001, Scarani and Gisin noticed that for $\sin2\xi \le 1/\sqrt{2^{N-1}}$
the states (\ref{eq1}) do not violate the
Mermin-Ardehali-Belinskii-Klyshko (MABK) inequalities. Based on
which, Scarani and Gisin wrote that ``this analysis suggests that
MK [in Ref. \cite{zukowski2}, MABK] inequalities, and more
generally the family of Bell's inequalities with two observables
per qubit, may not be the `natural' generalizations of the CHSH
inequality to more than two qubits" \cite{scarani}, where CHSH stands for Clauser-Horne-Shimony-Holt. In Ref.
\cite{zukowski1} {\.Z}ukowski and Brukner ({\.Z}B) have derived a general
Bell inequality for correlation
functions for $N$ qubits. The {\.Z}B inequalities include
MABK inequalities as special cases. Ref. \cite{zukowski2} shows
that (a) For $N={\rm even}$, although the generalized GHZ state
(\ref{eq1}) does not violate MABK inequalities, it violates the {\.Z}B
inequality and (b) For $\sin2\xi \le 1/\sqrt{2^{N-1}}$ and $N={\rm
odd}$, the correlations between measurements on qubits in the
generalized GHZ state (\ref{eq1}) satisfy all Bell inequalities
for correlation functions, which involve two dichotomic
observables per local measurement station.

In this Letter, we focus on a three-qubit system, whose corresponding
generalized GHZ state reads
$|\psi\rangle_{\rm GHZ}=\cos\xi|000\rangle+\sin\xi|111\rangle$. Up to now,
there is no Bell inequality violated by this pure entangled state
for the region $\xi \in (0, \pi/12]$ based on the standard Bell
experiment. Can Gisin's theorem be generalized to three-qubit pure
entangled states? Can one find a Bell inequality that violates
$|\psi \rangle_{\rm GHZ}$ for the whole region? In the following, we first present a
theorem that all generalized GHZ states of a three-qubit system
violate a Bell inequality in terms of probabilities; second, we
will provide a universal Bell inequality for probabilities
that is violated by all pure entangled states of a three-qubit system.

{\it Theorem 1: All generalized GHZ states of a three-qubit system
violate a Bell inequality for probabilities.}

{\it Proof:} Let us consider the following Bell-type scenario:
three space-separated observers, denoted by $A$, $B$ and $C$ (or
Alice, Bob and Charlie), can measure two different local
observables of two outcomes, labelled by 0 and 1. We denote
$X_i$ the observable measured by party $X$ and $x_i$ the outcome
with $X=A,B,C$ $(x=a,b,c)$. If the observers decide to measure
$A_1$, $B_1$ and $C_2$, the result is $(0,1,1)$ with probability
$P(a_1=0,b_1=1,c_2=1)$. The set of these $8\times 8$ probabilities
gives a complete description of any statistical quantity that can
be observed in this Gedanken experiment. One can easily see that,
any local-realistic (LR) description of the previous Gedanken
experiment satisfies the following Bell inequality:
\begin{widetext}
\begin{eqnarray}
&&P(a_1+b_1+c_1=0)+P(a_1+b_1+c_1=3)+P(a_1+b_2+c_2=2)  \nonumber \\
&&+P(a_2+b_1+c_1=0)+P(a_2+b_1+c_1=3)+P(a_2+b_2+c_2=1) \nonumber \\
&&-P(a_1+b_1+c_1=1)-P(a_1+b_2+c_2=1) \nonumber \\
&&-P(a_2+b_1+c_1=2)-P(a_2+b_2+c_2=2) \leq 2.
\label{bell1}
\end{eqnarray}
\end{widetext}
The joint probability $P(a_i+b_j+c_k=r)$, for instance,
$P(a_i+b_j+c_k=1)=P(a_i=1,b_j=0,c_k=0)+P(a_i=0,b_j=1,c_k=0)+P(a_i=0,b_j=0,c_k=1)$.
However, quantum mechanics will violate this Bell inequality for
any generalized GHZ states. The quantum prediction for the joint
probability reads
\begin{widetext}
\begin{eqnarray}
P^{QM}(a_i=m,b_j=n,c_k=l)= \langle \psi|
\hat{P}(a_i=m)\otimes\hat{P}(b_j=n)\otimes\hat{P}(c_k=l)
|\psi\rangle
 \label{eq2}
\end{eqnarray}
\end{widetext}
where $i, j, k=0,1$; $m,n,l=0,1$, and
\begin{eqnarray}
&&\hat{P}(a_i=m) =
\frac{1+(-1)^m\hat{n}_{a_i}\cdot\vec{\sigma}}{2} \nonumber\\
& & = \frac{1}{2}\left(\begin{array}{cc}
1+(-1)^m\cos\theta_{a_i} & (-1)^m\sin\theta_{a_i} e^{-i\phi_{a_i}} \\
(-1)^m\sin\theta_{a_i} e^{i\phi_{a_i}} &  1-(-1)^m\cos\theta_{a_i}
\end{array}     \right)
 \label{eq3}
\end{eqnarray}
is the projector of Alice for the $i$th measurement, and similar
definitions for $\hat{P}(b_j=n)$, $\hat{P}(c_k=l)$. More
precisely, for the generalized GHZ state one obtains

\begin{widetext}
\begin{eqnarray}
&&P^{QM}(a_i=m,b_j=n,c_k=l)\nonumber\\
&&=\frac{1}{8}\cos^2\xi[1+(-1)^m\cos\theta_{a_i}][1+(-1)^n\cos\theta_{b_j}][1+(-1)^l\cos\theta_{c_k}]\nonumber\\
 &&+\frac{1}{8}\sin^2\xi[1-(-1)^m\cos\theta_{a_i}][1-(-1)^n\cos\theta_{b_j}][1-(-1)^l\cos\theta_{c_k})]\nonumber \\
&&+\frac{1}{8}\sin(2\xi)
(-1)^{m+n+l}\sin\theta_{a_i}\sin\theta_{b_j}\sin\theta_{c_k}\cos(\phi_{a_i}+\phi_{b_j}+\phi_{c_k}).
\end{eqnarray}
\end{widetext}
For convenient reason, let us denote the left-hand side of the Bell inequality by ${\cal B}$, which represents a Bell quantity.
For the following settings
$\theta_{a_1}=\theta_{a_2}=\theta,\phi_{a_1}=-\pi/3,
\phi_{a_2}=2\pi/3, \theta_{b_1}=\theta_{c_1}=0,
\phi_{b_1}=\phi_{c_1}=0, \theta_{b_2}=
\theta_{c_2}=\pi/2, \phi_{b_2}=\phi_{c_2}=\pi/6$, the Bell
quantity ${\cal B}$ is given as
${\cal B}  = \frac{1}{2}+\frac{3}{2}(\cos\theta+\sin(2\xi)\sin\theta)
\ge \frac{1}{2}+\frac{3}{2}\sqrt{1+\sin^2(2\xi)}$,
the equal sign occurs at $\theta=\tan^{-1}[\sin(2\xi)]$. Obviously the
Bell inequality is violated for any $\xi\ne 0$ or $\pi/2$. This ends the proof.

This theorem indicates that it is possible for a Bell inequality
in terms of probabilities to be violated by all pure entangled
generalized GHZ states. Recently, classification of $N$-qubit
entanglement via a quadratic Bell inequality consisting of MABK
polynomials has been presented in Ref. \cite{yu}. For $N=3$, there are
three types of three-qubit states: i) totally separable states denoted
as $(1_3)$=\{ mixtures of states of form
$\rho_A\otimes\rho_B\otimes \rho_C$\}; ii) 2-entangled states
which are denoted as $(2,1)$=\{mixtures of states of form
$\rho_A\otimes \rho_{BC}, \rho_{AC}\otimes\rho_B, \rho_{AB}\otimes
\rho_C$\}; iii) fully entangled states which are denoted as
$(3)=\{\rho_{ABC}\}$ including the GHZ state. Ref. \cite{yu} has
drawn an ancient Chinese coin (ACC) diagram for the classification
of three-qubit entanglement. However, for the four points located on
the four corners of the square, some of the above three types of
three-qubit states coexist. For instance, the totally separable states
and the generalized GHZ states for $\xi\in (0,\pi/12]$ coexist at
these four corners, it looks somehow like these four points are
``degenerate''. The above Bell inequality for probabilities is
useful, at least; it can distinguish the generalized GHZ states
for $\xi\in (0,\pi/12]$ from the totally separable states.

There are two different entanglement classes for three-qubit states,
namely, 2-entangled states and fully entangled states. Why MABK
inequalities as well as {\.Z}B inequalities fail for the region $\xi \in
(0, \pi/12]$ may be due to the reason that their inequalities
contain only fully three-particle correlations. If one expands
$\hat{P}(a_i=m)\otimes\hat{P}(b_j=n)\otimes\hat{P}(c_k=l)$ and
substitutes them into a Bell quantity ${\cal B}$, one will find
that ${\cal B}$ contains not only the terms of fully three-particle correlations,
such as $\hat{n}_{a_i}\cdot\vec{\sigma} \otimes
\hat{n}_{b_j}\cdot\vec{\sigma}\otimes \hat{n}_{c_l}\cdot\vec{\sigma}$, but also
the terms of two-particle correlations, such as
$\hat{n}_{a_i}\cdot\vec{\sigma} \otimes \hat{n}_{b_j}\cdot\vec{\sigma}\otimes {\bf
1}$. The above theorem implies that two-particle correlations may
make a contribution to the quantum violation of Bell inequality. The remarkable property of the Bell inequality in Eq. (\ref{bell1}) is that it is violated by all pure entangled generalized GHZ states. However, some of other pure entangled states does
not violate it, such as the W state $|\psi \rangle_W = (|100\rangle +|010\rangle +|001\rangle)/\sqrt{3}$. The reason
may be that the Bell inequality in Eq. (\ref{bell1}) does not contain all the possible probabilities. This motivates us to introduce a Bell inequality
with all possible probabilities:
\begin{widetext}
\begin{eqnarray}
&& P(a_1+b_1+c_1=1)+ 2 P(a_2+b_2+c_2=1)  \nonumber \\
&& +P(a_1+b_2+c_2=2)+P(a_2+b_1+c_2=2)+P(a_2+b_2+c_1=2) \nonumber \\
&& -P(a_1+b_1+c_2=0)-P(a_1+b_2+c_1=0)-P(a_2+b_1+c_1=0) \nonumber \\
&& -P(a_1+b_1+c_2=3)-P(a_1+b_2+c_1=3)-P(a_2+b_1+c_1=3) \leq 3.
\label{bell2}
\end{eqnarray}
\end{widetext}
This inequality is symmetric under the permutations of three observers: Alice, Bob and Charlie. Pure states of three qubits constitute a five-parameter family, with equivalence up to local unitary transformations. This family has the representation \cite{acin}
\begin{eqnarray}
|\psi\rangle & = & \sqrt{\mu_0}|000\rangle+\sqrt{\mu_1}e^{i\phi}|100\rangle
+\sqrt{\mu_2}|101\rangle \nonumber\\
&& +\sqrt{\mu_3}|110\rangle+\sqrt{\mu_4}|111\rangle
 \label{state}
\end{eqnarray}
with $\mu_i \ge 0$, $\sum_i \mu_i=1$ and $0 \le \phi \le \pi$. Numerical results show that this Bell
inequality for probabilities is violated by all pure entangled states of a three-qubit system. However, it is difficult to provide an analytic proof.
\begin{figure}
\begin{center}
\epsfig{figure=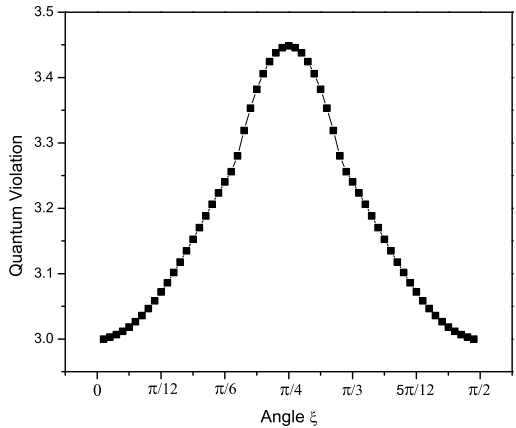,width=0.45\textwidth}
\end{center}
\caption{ Numerical results for the generalized GHZ states $|\psi\rangle_{\rm GHZ}=\cos\xi|000\rangle+\sin\xi|111\rangle$,
which violate a Bell inequality for probabilities (\ref{bell2}) except $\xi=0$ and $\pi/2$. For the GHZ state with $\xi=\pi/4$, the Bell
quantity reaches its maximum value $\frac{3}{8}(4+3\sqrt{3})$.} \label{fig1}
\end{figure}

\begin{figure}
\begin{center}
\epsfig{figure=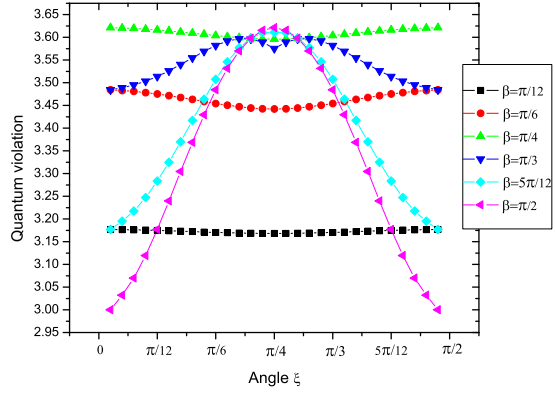,width=0.45\textwidth}
\end{center}
\caption{Numerical results for the family of generalized W states $|\psi\rangle_{W}=\sin\beta\cos\xi|100\rangle+\sin\beta\sin\xi|010\rangle+\cos\beta|001\rangle$
with the cases $\beta=\pi/12, \pi/6,\pi/4,\pi/3,5\pi/12$ and $\pi/2$. } \label{fig2}
\end{figure}

In Fig. \ref{fig1}, we show the numerical results for the generalized GHZ
states $|\psi\rangle_{\rm GHZ}=\cos\xi|000\rangle+\sin\xi|111\rangle$,
which violate the above symmetric Bell inequality for
probabilities except $\xi=0$ and $\pi/2$. For the measuring angles
$\theta_{a_1}=\theta_{a_2}=\theta_{b_1}=\theta_{b_2}=\theta_{c_1}=\theta_{c_2}=\pi/2,
\phi_{a_1}=-5\pi/12, \phi_{a_2}=\pi/4, \phi_{b_1}=-5\pi/12,
\phi_{b_2}=\pi/4, \phi_{c_1}=-\pi/3, \phi_{c_2}=\pi/3$, all the
probability terms with positive signs in the Bell inequality
(\ref{bell2}) are equal to $\frac{3}{16}(2+\sqrt{3})$, while the
terms with negative signs are equal to $\frac{1}{8}$, so the
quantum violation of the Bell quantity for the GHZ state (where
$\xi=\pi/4$) is obtained as
$6\times\frac{3}{16}(2+\sqrt{3})-6\times\frac{1}{8}=\frac{3}{8}(4+3\sqrt{3})>3$.
In Fig. \ref{fig2}, we show the numerical results for the family of
generalized W states
$|\psi\rangle_{W}=\sin\beta\cos\xi|100\rangle+\sin\beta\sin\xi|010\rangle+\cos\beta|001\rangle$
with the cases $\beta=\pi/12, \pi/6,\pi/4,\pi/3,5\pi/12$ and
$\pi/2$, which show the quantum violation of $|\psi\rangle_W$
except the product cases with $\beta=\pi/2, \xi=0$ and
$\pi/2$. For the standard W state
$|\psi\rangle_{W}=(|100\rangle+|010\rangle+|001\rangle)/\sqrt{3}$,
the quantum violation is $3.55153$. We now proceed to present the second theorem.

{\it Theorem 2: All pure 2-entangled states of a three-qubit system violate a Bell
inequality for probabilities.}

{\it Proof:} By pure 2-entangled states of three-qubit system, we
mean $|\psi_{AB}\rangle\otimes|\psi_C\rangle$,
$|\psi_{AC}\rangle\otimes|\psi_B\rangle$ and
$|\psi_{BC}\rangle\otimes|\psi_A\rangle$. It is sufficient to
consider one of them, say
$|\psi_{AB}\rangle\otimes|\psi_C\rangle$, since the Bell inequality
(\ref{bell2}) is symmetric under the permutations of $A$, $B$ and
$C$. Moreover, one can always have
$|\psi_{AB}\rangle\otimes|\psi_C\rangle= (\cos\xi|00\rangle_{AB}
+\sin\xi|11\rangle_{AB})\otimes |0\rangle_C$ due to local unitary
transformations. For the measuring angles
$\theta_{a_1}=\theta_{a_2}=\theta,\phi_{a_1}=2\pi/3,
\phi_{a_2}=-\pi/3, \theta_{b_1}=\theta_{c_1}=0,
\phi_{b_1}=\phi_{c_1}=0, \theta_{b_2}=\pi/2, \theta_{c_2}=\pi,
\phi_{b_2}=\pi/3,\phi_{c_2}=0$, we obtain from the left-hand side
of the Bell inequality (\ref{bell2}) that
${\cal B}  =\frac{3}{2}(1-\cos\theta+\sin(2\xi)\sin\theta)
\ge \frac{3}{2}(1+\sqrt{1+\sin^2(2\xi)})$,
the equal sign occurs at $\theta=-\tan^{-1}[\sin(2\xi)]$. Obviously the
Bell inequality is violated for any $\xi\ne 0$ or $\pi/2$. This ends the proof. 
Indeed, the quantum violation of the state $|\psi_{AB}\rangle\otimes|\psi_C\rangle$
corresponds to the curve with $\beta=\pi/2$ as shown in Fig.2, because $|\psi_{AB}\rangle\otimes|\psi_C\rangle$ is equivalent to $|\psi\rangle_W$ for $\beta=\pi/2$
up to a local unitary transformation.

There is a simpler and more intuitive way to prove Theorem 2, because the symmetric Bell inequality (\ref{bell2}) can be reduced to a CHSH-like inequality for two qubits and then from Gisin's theorem for two qubits, one easily has Theorem 2. By 
taking $c_1=0, c_2=1$, we have from Eq. (\ref{bell2}) that
\begin{eqnarray}
&& P(a_1+b_1=1)+ 2 P(a_2+b_2=0)  \nonumber \\
&& +P(a_1+b_2=1)+P(a_2+b_1=1)+P(a_2+b_2=2) \nonumber \\
&& -P(a_1+b_1=-1)-P(a_1+b_2=0)-P(a_2+b_1=0) \nonumber \\
&& -P(a_1+b_1=2)-P(a_1+b_2=3)-P(a_2+b_1=3) \leq 3
\end{eqnarray}
Since $a_1,a_2, b_1,b_2=0,1$, the probabilities $P(a_1+b_1=-1)$, $P(a_1+b_2=3)$ and$P(a_2+b_1=3)$ will be equal to zero, by 
using $P(a_2+b_2=0)+P(a_2+b_2=2)=1-P(a_2+b_2=1)$, we arrive at the following Bell inequality for two-qubit:
\begin{widetext}
\begin{eqnarray}
&& P(a_1+b_1=1)+ P(a_1+b_2=1)+ P(a_2+b_1=1)+ P(a_2+b_2=0)  \nonumber \\
&& -P(a_1+b_1=2)- P(a_1+b_2=0)- P(a_2+b_1=0)- P(a_2+b_2=1) \leq 2.
\label{bell3}
\end{eqnarray}
\end{widetext}
This Bell inequality is symmetric under the permutations of Alice
and Bob, it is an alternative form for CHSH inequality of two qubits. For the two-qubit state
$|\psi\rangle=\cos\xi|00\rangle+\sin\xi|11\rangle$ and the
projector as shown in Eq.(\ref{eq3}), one can have the quantum
probability
$P^{QM}(a_i=m,b_j=n)
=\frac{1}{4}\cos^2\xi[1+(-1)^m\cos\theta_{a_i}][1+(-1)^n\cos\theta_{b_j}]
+\frac{1}{4}\sin^2\xi[1-(-1)^m\cos\theta_{a_i}][1-(-1)^n\cos\theta_{b_j}]
+\frac{1}{4}\sin(2\xi)
(-1)^{m+n}\sin\theta_{a_i}\sin\theta_{b_j}\cos(\phi_{a_i}+\phi_{b_j})
$.
For the measuring angles
$\theta_{a_1}=\theta_{a_2}=\theta,\phi_{a_1}=\pi-\phi, \phi_{a_2}=-\phi, \theta_{b_1}=0,
\phi_{b_1}=0, \theta_{b_2}=\pi/2, \phi_{b_2}=\phi$, the left-hand side of a Bell inequality (\ref{bell3}) becomes ${\cal B}=\frac{1}{2}+\frac{3}{2}(-\cos\theta+\sin(2\xi)\sin\theta)\ge \frac{1}{2}(1+3\sqrt{1+\sin^2(2\xi)})$, the equal sign occurs at $\theta=-\tan^{-1}[\sin(2\xi)]$. Obviously a Bell inequality (\ref{bell3}) is violated for any $\xi\ne 0$ or $\frac{\pi}{2}$, just the same as CHSH inequality violated by the two-qubit state $|\psi\rangle=\cos\xi|00\rangle+\sin\xi|11\rangle$.
For the Werner state $\rho_W=V|\psi\rangle\langle\psi|+(1-V)\rho_{\rm noise}$, where $|\psi\rangle=(|00\rangle +|11\rangle)/\sqrt{2}$ is the maximally entangled state. The maximal value of $V$ that a local realism is still possible by this Bell inequality is $V_{\rm max}=1/\sqrt{2}$, just the same as 
the case for CHSH inequality. Actually, if one denotes the left-hand side of Bell inequality (\ref{bell3}) by ${\cal B}$ and redefines a new Bell quantity ${\cal B}'= \frac{4}{3}({\cal B}-\frac{1}{2})$, he still has a Bell inequality
${\cal B}' \le 2$. For quantum mechanics, ${\cal B}' _{\rm max}=2 \sqrt{1+\sin^2(2\xi)}$, which reaches $2\sqrt{2}$ and then ${\cal B}'$ recovers the usual CHSH inequality. 

Theorem 2 is remarkable. If one knows that a pure state is a 2-entangled state of a three-qubit system,
one can use a Bell inequality (\ref{bell2}) to measure the degree of entanglement (or concurrence denoted by
${\cal C}$) of the state. Since ${\cal B}_{\rm max} = \frac{3}{2}(1+\sqrt{1+\sin^2(2\xi)}) = \frac{3}{2}(1+\sqrt{1+{\cal C}^2})$, 
thus one has the concurrence
${\cal C}=|\sin(2\xi)| \in [0,1]$, just the same as the case of CHSH inequality measure of the concurrence of pure states
of two qubits. In summary, (i) since all pure entangled states (including pure 2-entangled states) of a three-qubit system violate a Bell inequality (\ref{bell2}), thus we have Gisin's theorem for a three-qubit system; (ii) the Bell inequality (\ref{bell2}) can be reduced to an alternative form of 
the CHSH inequality (in terms of probabilities), thus it can be viewed as a good candidate for a ``natural" generalization of the usual CHSH inequality. (iii) MABK inequalities and {\.Z}B
 inequalities are binary correlation Bell inequalities. However, one may notice that Bell inequalities (\ref{bell1}) and (\ref{bell2}) are both ternary Bell inequalities, i.e., where the inequalities are ``modulo 3". Most recently, a ternary Bell inequality in terms of probabilities for three qutrits was presented in Ref. \cite{acin2}, this inequality can be connected to the Bell inequality (\ref{bell2}), which is for three qubits if one restricts the initial three possible outcomes of each measurement to only two possible outcomes.

We thank M. \.{Z}ukowski for valuable discussion. This work is supported by NUS academic research Grant No. WBS: R-144-000-089-112. J.L. Chen acknowledges financial support from Singapore Millennium Foundation and (in part) by NSF of China (Grant No. 10201015). C. Wu acknowledges financial support by NUS.

\end{document}